%% file: arxiv.tex
\documentclass[11pt]{article}

\usepackage[margin=1in]{geometry}

\usepackage{amsmath,amssymb,amsthm}
\usepackage{graphicx}
\usepackage{booktabs}
\usepackage{array,tabularx}
\usepackage{float}
\usepackage{xcolor}

\usepackage[numbers,sort&compress]{natbib}
\usepackage[colorlinks=true, citecolor=green!60!black, linkcolor=blue!60!black]{hyperref}

\newcommand{\xhdr}[1]{\vspace{0mm}\noindent{{\bf #1.}}}
\newcolumntype{L}[1]{>{\raggedright\arraybackslash}p{#1}}
\newcolumntype{Y}{>{\raggedright\arraybackslash}X}

\theoremstyle{plain}

\theoremstyle{definition}

\theoremstyle{remark}

\title{Do LLMs Track Public Opinion?\\
A Multi-Model Study of Favorability Predictions in the 2024 U.S. Presidential Election}

\author{
Riya Parikh \\
MIT ORC \\
\texttt{riyapar@mit.edu} \and 
Sarah H. Cen \\
CMU \\ 
\texttt{sarahcen@andrew.cmu.edu} \and 
Chara Podimata \\
MIT \\
\texttt{podimata@mit.edu}
}

\date{}

\begin{document}
\maketitle

\begin{abstract}
We investigate whether Large Language Models (LLMs) can track public opinion as measured by exit polls during the 2024 U.S. presidential election cycle. Our analysis focuses on headline favorability (e.g., ``Favorable'' vs.\ ``Unfavorable'') of presidential candidates across multiple LLMs queried daily throughout the election season. Using the publicly available \texttt{llm-election-data-2024} dataset, we evaluate predictions from nine LLM configurations against a curated set of five high-quality polls from major organizations (Reuters, CNN, Gallup, Quinnipiac, ABC). We find systematic directional miscalibration. For Kamala Harris, all models overpredict favorability by 10--40\% relative to polls. For Donald Trump, biases are smaller (5--10\%) and poll-dependent, with substantially lower cross-model variation. These deviations persist under temporal smoothing and are not corrected by internet-augmented retrieval. We conclude that off-the-shelf LLMs do not reliably track polls when queried in a straightforward manner and discuss implications for election forecasting.
\end{abstract}

\input{intro}

\input{relatedwork}
\input{methodology}

\input{results}

\input{discussion}

\newpage
\bibliographystyle{unsrtnat}
\bibliography{references}

\appendix
\onecolumn
\section{Supplementary Materials}

\subsection{Kamala Harris Supplementary Plots}
\label{sec:Kamala-supplementary}

\begin{figure}[H]
    \centering
    \includegraphics[width=0.7\linewidth]{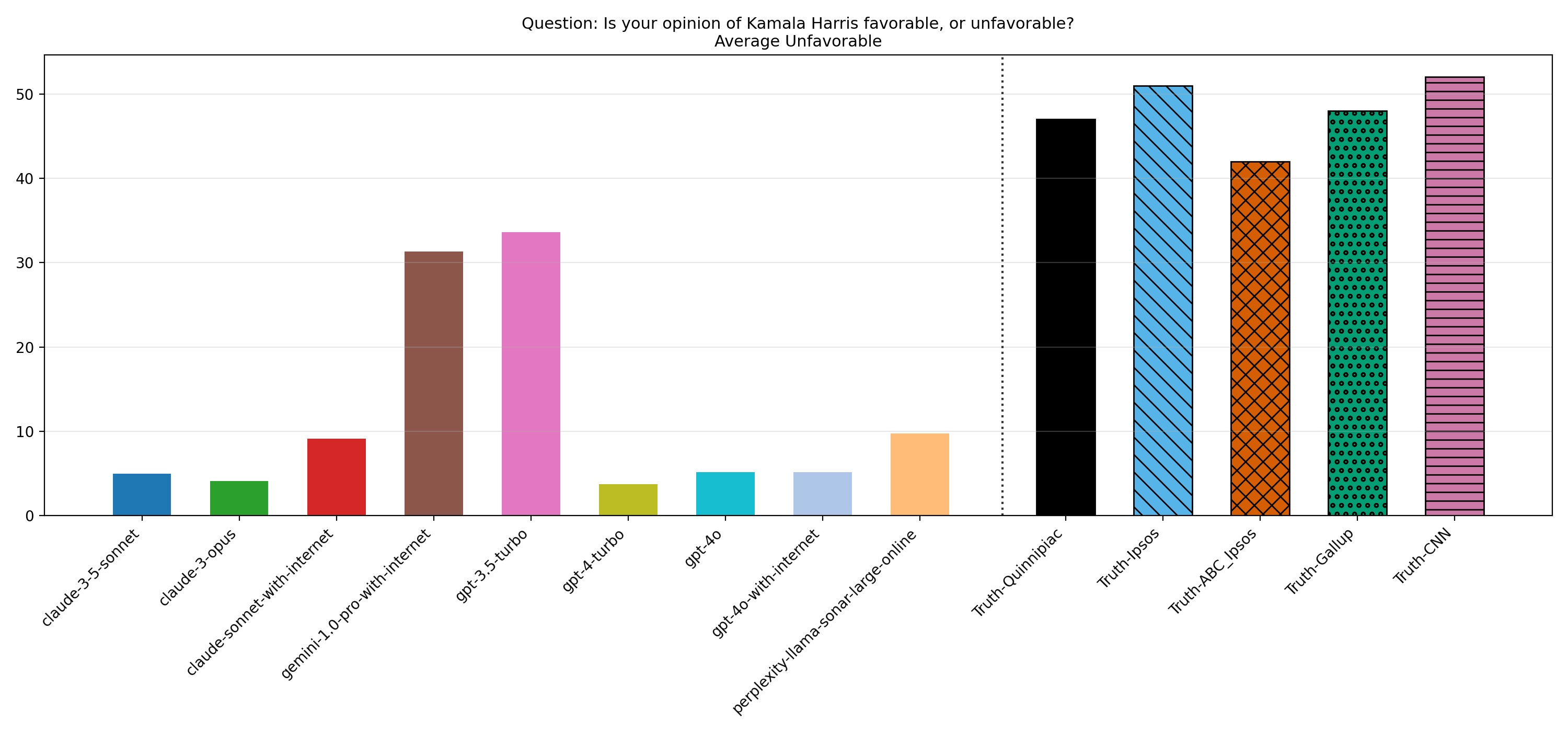}
    \caption{Average-over-time predictions of Kamala Harris unfavorability (9 left bars) vs actual polls (5 right bars).}
    \label{fig:kamala-unfavorable-barplot}
\end{figure}

\begin{figure}[H]
    \centering
    \includegraphics[width=0.7\linewidth]{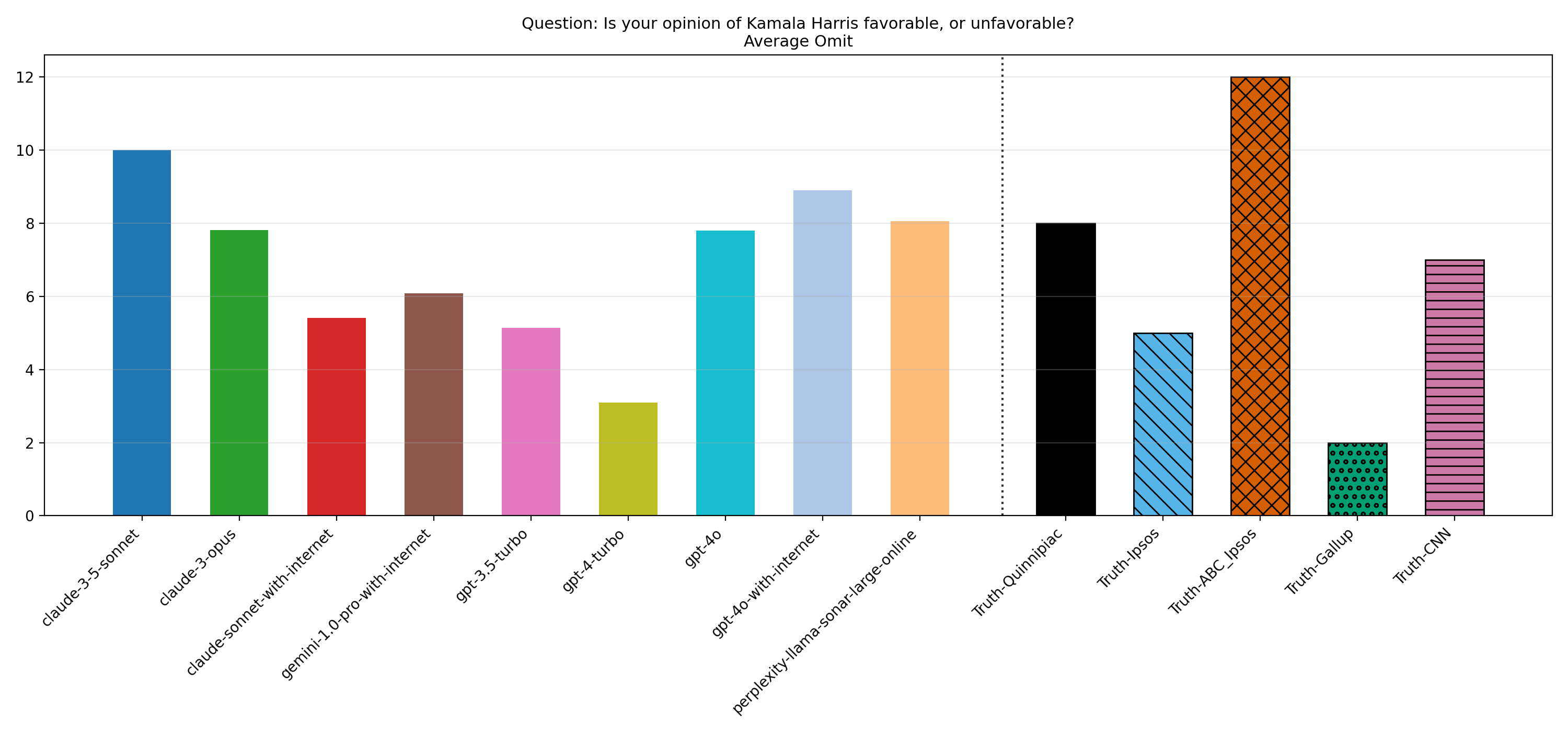}
    \caption{Average-over-time predictions of ``Omit'' for Kamala Harris (9 left bars) vs actual polls (5 right bars).}
    \label{fig:kamala-omit-barplot}
\end{figure}

\begin{figure}[H]
    \centering
    \includegraphics[width=0.7\linewidth]{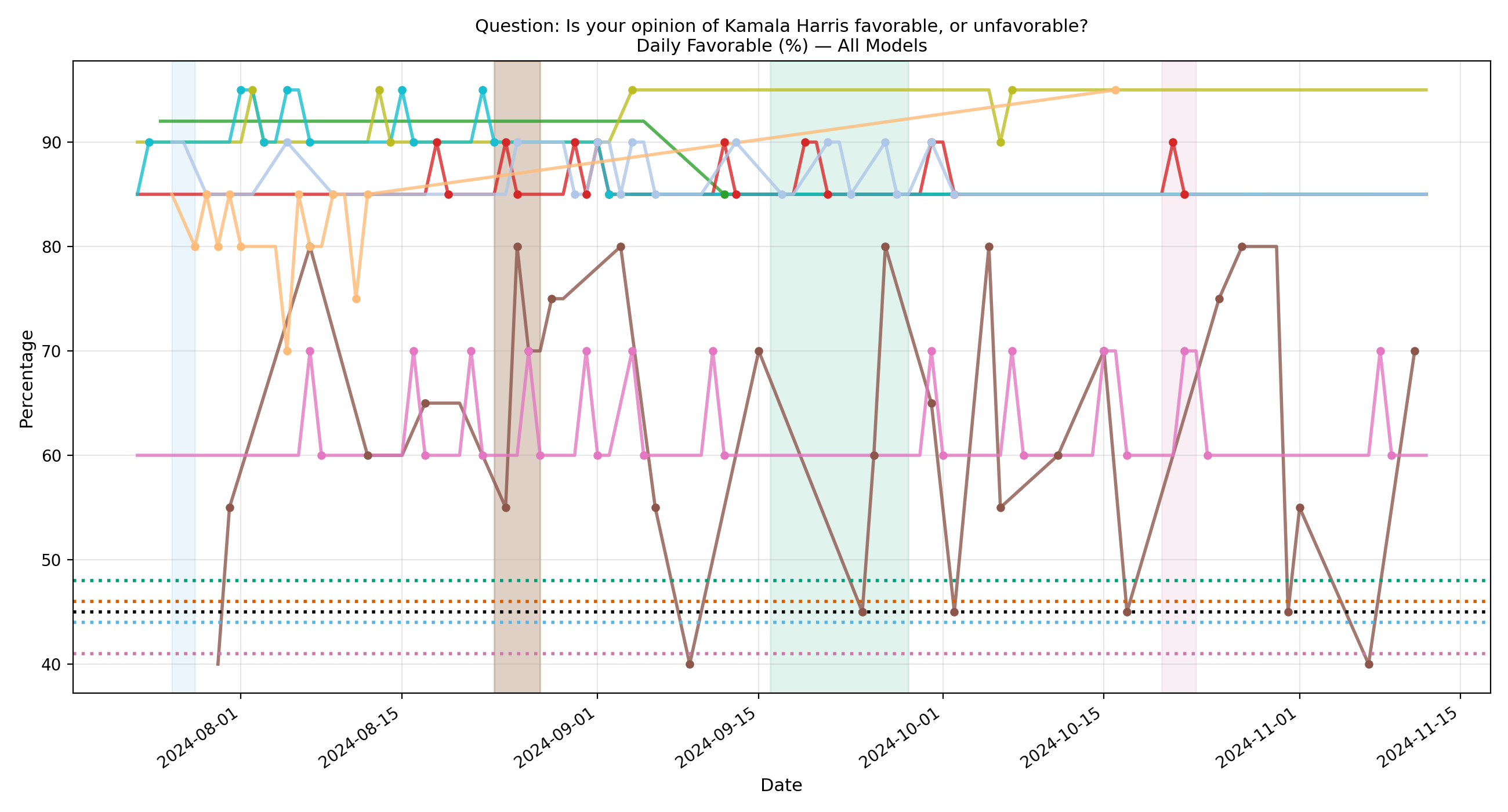}
    \caption{Daily responses on Kamala Harris favorability predictions. Solid lines: LLM outputs; dashed lines: ground-truth polls. Shaded windows indicate poll field periods.}
    \label{fig:daily-kamala-favorability}
\end{figure}

\begin{figure}[H]
    \centering
    \includegraphics[width=0.7\linewidth]{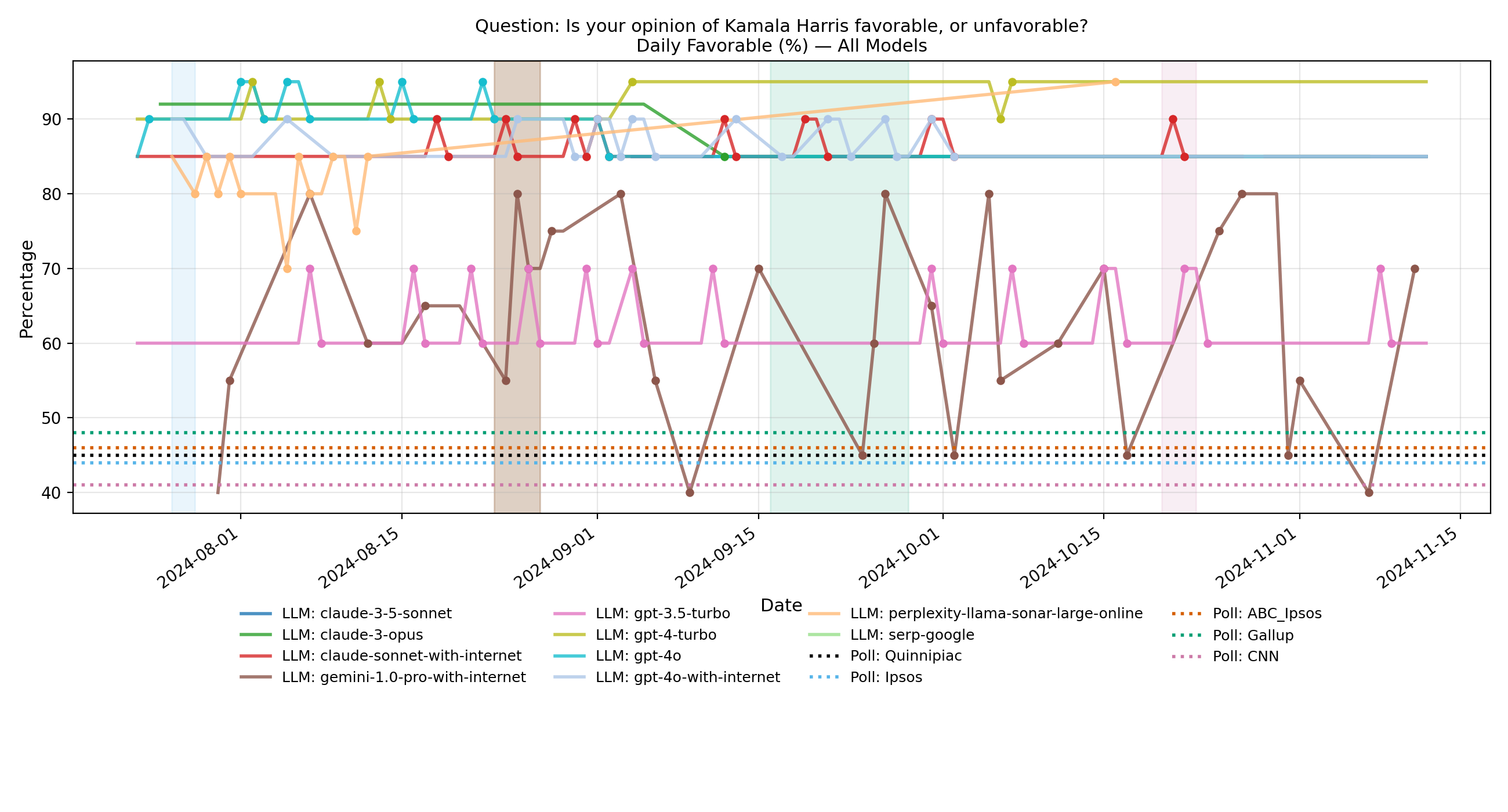}
    \caption{Daily response change events for Kamala Harris favorability predictions across models.}
    \label{fig:daily-changes-kamala-favorability}
\end{figure}

\subsection{Donald Trump Supplementary Plots}
\label{sec:trump-supplementary}

\begin{figure}[H]
    \centering
    \includegraphics[width=0.7\linewidth]{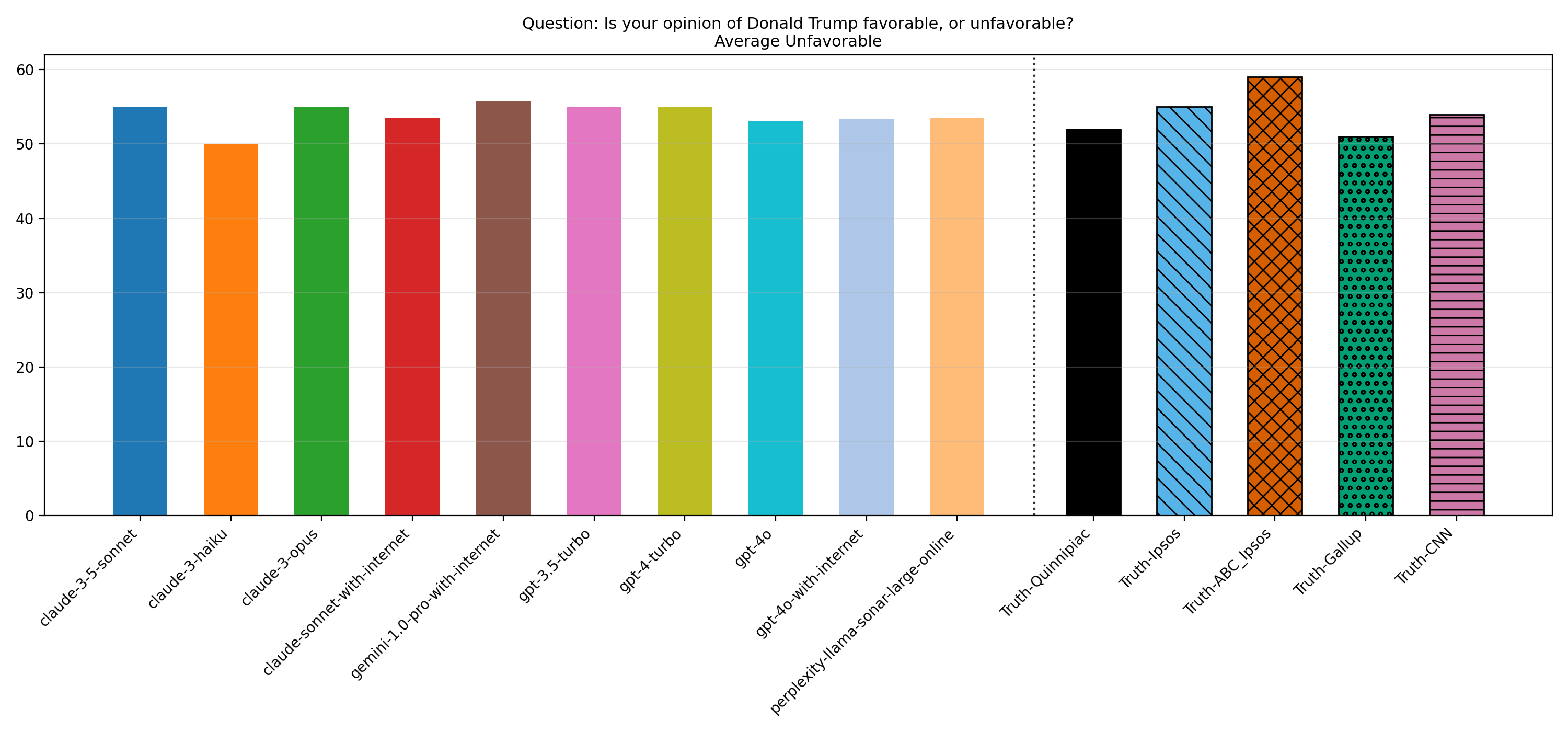}
    \caption{Average-over-time predictions of Donald Trump unfavorability (9 left bars) vs actual polls (5 right bars).}
    \label{fig:trump-unfavorable-barplot}
\end{figure}

\begin{figure}[H]
    \centering
    \includegraphics[width=0.7\linewidth]{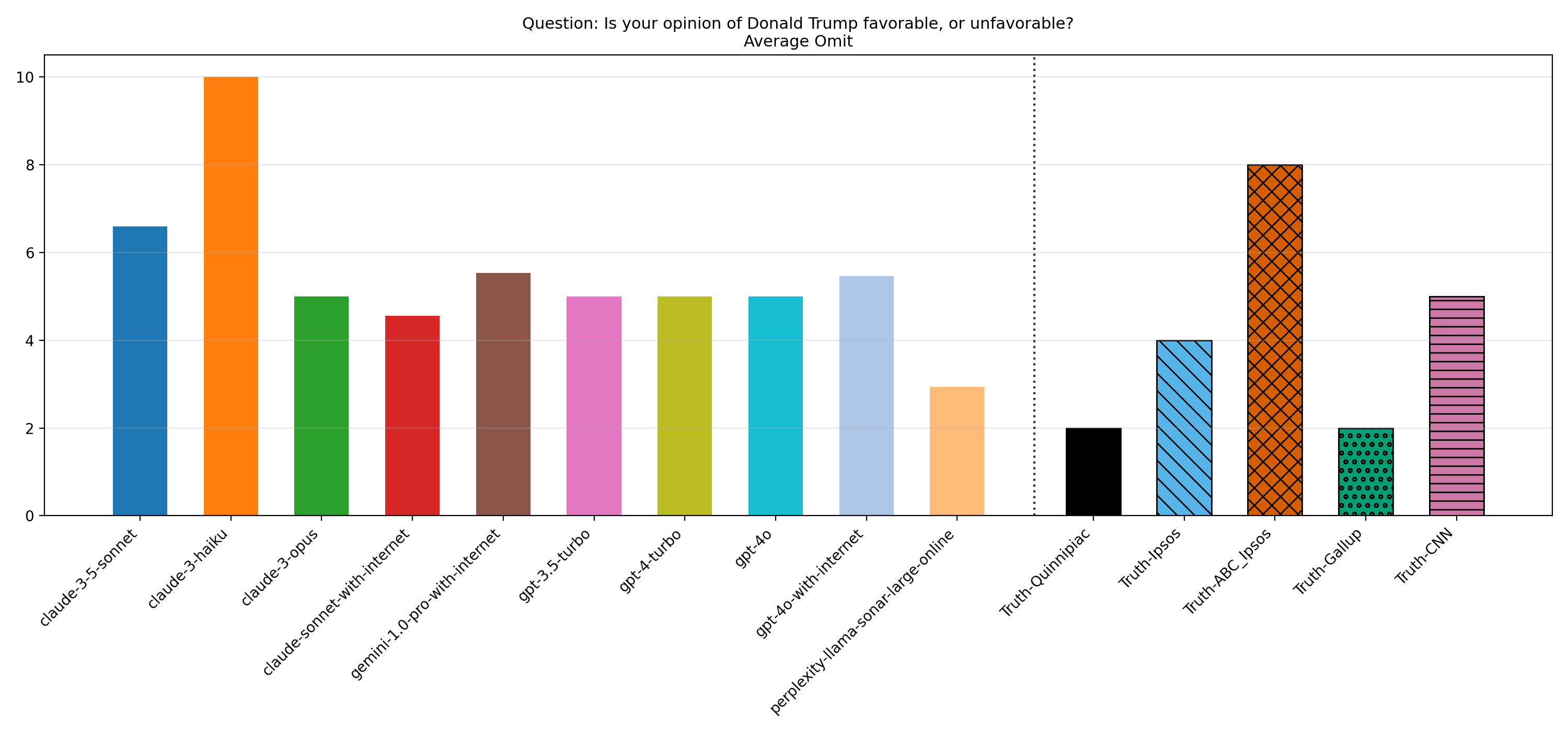}
    \caption{Average-over-time predictions of ``Omit'' for Donald Trump (9 left bars) vs actual polls (5 right bars).}
    \label{fig:trump-omit-barplot}
\end{figure}

\end{document}

%% file: intro.tex
\section{Introduction}\label{sec:intro}


Traditional polling faces mounting challenges~\cite{pew_low_response_2017}. Response rates to telephone surveys have declined dramatically from over 35\% in the 1990s to 6\% in 2018~\citep{kennedy_hartig_2019}, raising concerns about nonresponse bias and the representativeness of survey samples. High-profile polling misses in recent U.S. elections~\cite{kennedy2018evaluation,pew_2020_poll_errors_2021}, including underestimates of support for Donald Trump in 2016 and 2020, have fueled public skepticism about the reliability of polls. Yet accurate polling remains essential: election forecasters depend on polls to model voting dynamics and estimate win probabilities, while policymakers and campaigns use polling data to gauge public sentiment on policy platforms and adjust their strategies accordingly. The challenge, thcsen, is not whether we need polling (we do!), but how to make it more accurate and cost-effective in an era when traditional methods are increasingly strained.


Large language models (LLMs) have been proposed as a potential solution to these polling challenges  \citep{berger_schneier_gong_sanders_2024}. Trained on vast corpora of text that include news articles, social media posts, opinion pieces, and other public discourse, LLMs may implicitly encode patterns of public sentiment and belief. Crucially, unlike traditional polls which must actively recruit respondents who are increasingly unwilling to participate, LLMs can draw on the billions of expressions of political opinion that people voluntarily share online, from social media posts to comment sections to forum discussions. This gives LLMs potential access to a far broader and potentially more representative sample of genuine public sentiment than pollsters can reach through phone calls or survey panels. Moreover, LLMs with internet access can continuously incorporate new information as events unfold, potentially tracking shifts in public opinion in real time rather than waiting for the next polling cycle. This has led researchers to ask whether LLMs can be queried to simulate or predict population-level opinions, effectively functioning as a kind of \emph{synthetic polling mechanism}~\citep{berger_schneier_gong_sanders_2024,schneier_sanders_2025}. Prior work in political science and computational social science has explored this idea primarily through the lens of \emph{synthetic personas}: researchers prompt LLMs to adopt specific demographic or partisan identities (e.g., ``You are a 45-year-old Republican woman from Ohio'') and then elicit responses to survey questions, treating the aggregated outputs as a proxy for how real subpopulations might answer~\cite{argyle2023out}. This approach has shown promise for understanding issue-based attitudes and has been studied across various policy domains~\cite{moon2024virtual,jiang2024donald}.


In this work, we focus on predictions that are suggestive of the election outcome. 
That is, while most existing work using LLMs for opinion measurement focus on issue positions, we examine LLMs' abilities to forecast the election.
There is growing interest in whether LLMs can directly forecast electoral outcomes---or at least track the kinds of summary statistics (like candidate favorability) that are central inputs to election forecasting models. Unlike issue polling, which measures attitudes on relatively stable topics, election forecasting requires models to capture rapidly evolving public sentiment about specific candidates during a compressed time window. If LLMs could reliably track such sentiment, they might serve as a faster and cheaper complement to traditional polling infrastructure.





The potential use of LLMs as polling proxies should be understood in the context of the broader election forecasting ecosystem. Traditional forecasting models (e.g., FiveThirtyEight, or the New York Times needle) aggregate multiple polls over time, weight them by pollster quality, recency, and other internal factors, and finally combine them with structural factors like economic indicators and incumbency effects. More recently, prediction markets like Kalshi and PredictIt have emerged as alternative signals of electoral outcomes, with market prices reflecting the collective beliefs of traders willing to stake money on election results. LLMs could potentially serve a complementary role: they are cheaper and faster than conducting new polls, and unlike prediction markets, they do not require liquid trading or financial stakes. The ultimate hope would be to determine whether LLM-based predictions could augment or validate traditional polling averages, particularly in contexts where polls are sparse, costly, or subject to high uncertainty. However, before LLMs can be integrated into forecasting pipelines, we need rigorous empirical evaluation of how well they actually track public opinion.

In this paper, we study a concrete version of this question: \emph{Do LLMs track public opinion as measured by traditional exit polls?} Specifically, we focus on candidate favorability questions for Kamala Harris and Donald Trump during the 2024 U.S. presidential election cycle. We leverage the publicly available \texttt{llm-election-data-2024} dataset~\footnote{\url{https://huggingface.co/datasets/sarahcen/llm-election-data-2024}} by~\citet{cen2025longitudinal}, which contains daily queries to 11 different LLM configurations, and we compare their predictions to a carefully selected sample of five high-quality polls from major organizations (Reuters, CNN, Gallup, Quinnipiac, and ABC). 
Part of this dataset includes LLMs' daily predictions of exit polling, 
where each LLM is queried to predict the percentage of voters that will select each of the multiple-choice options of anticipated exit poll questions.\footnote{Exit poll questions are not released before the election. \citet{cen2025longitudinal} state that they use past exit polls to construct questions that will likely be asked.}

Our methodology is as follows: we parse the LLM predictions on candidate favorability (``Favorable'', ``Unfavorable'', ``Omit'') per candidate.
We subsequently align these distributions with favorability questions of real polls to assess whether LLMs track polling across time faithfully. B
Although the dataset does contain LLM responses under various prompt variations (e.g., demographic steering and polling by voter groups), we restrict our attention to ``general population'' prompts (i.e., queries without demographic, partisan or other conditioning). 
Doing so allows us to ask whether off-the-shelf LLMs can approximate the aggregate electorate's favorability ratings, not just subgroup opinions. Our findings reveal systematic directional miscalibration: Harris' favorability is consistently overpredicted (and unfavorability underpredicted) across models. The situation is less clear for Trump's favorability; models some times over- and other times under-predict his favorability compared to the ground truth polls, thus leading to inconsistent results. Additionally, there seems to be more cross-model variation for Harris' favorability versus Trump's. 

The rest of the paper is organized as follows: Section~\ref{sec:related-work} includes background and related work. Section~\ref{sec:methodology} explains our methodology. Section~\ref{sec:results} documents our results. We conclude our paper with an extensive discussion on the future directions of this research agenda in Section~\ref{sec:discussion}.

%% file: relatedwork.tex
\section{Related Work}
\label{sec:related-work}

\paragraph{LLMs for human subject research.}
Using LLMs to emulate human survey responses has been an area of longstanding interest. 
Early works, such as \citep{santurkar2023whose}, study the extent to which LLMs reflect the opinions of different demographic groups using Pew surveys as a baseline, finding that models perform well for some populations but deviate for others. 
A series of works also study the ability of LLMs to simulate human behavior in a variety of settings. 
\citet{park2023generative} examines the use of generative AI agents to produce simulacra---or representations---of humans in a sandbox; \cite{park2024generative} scale this effort to the simulations of 1,000 agents, demonstrating large-scale social simulation and behavior modeling is possible with LLM-based agents.
Other works focus specifically on the promise of LLMs for social science research and surveys, such as \citep{argyle2023out}. 
For example, 
\citet{moon2024virtual} explore the ability of LLMs to emulate more diverse survey respondents by feeding LLMs ``backstories.''  
\citet{kolluri2025finetuning} use fine-tuning to more closely align LLM behavior with human behavior in the context of social science behavior. 
Still, many are wary of using LLMs to simulate humans, as they may not capture important nuances or deviate towards the mean.
For example, synthetic survey data can misrepresent or flatten demographic variation \cite{bisbee2024synthetic,wang2024replace}, and using LLMs to simulate behavior raises causal and methodological challenges \cite{gui2023challenge}.
There is an extensive literature that we refer the reader to for further work \cite{park2022social,horton2023agents,tjuatja2024survey,karanjai2025rolecreation,cao2025specialize}.

\paragraph{LLMs for forecasting.}
LLMs have also been applied to forecasting, particularly of time series data.
Within time series forecasting, methods include reprogramming pretrained LLMs for numerical series that outperform other specialized models \cite{jin2023time}, autoregressive forecasters built on LLMs that use in-context forecasting and exhibit significant speedups \cite{liu2024autotimes}, and integration of event or news analysis with reflection to improve accuracy \cite{wang2024news}.
Some compare LLMs against standard statistical models \cite{cao2024evaluation} and ask whether they are particularly useful for forecasting \cite{tan2024language}.
\citet{halawi2024approaching} develop a retrieval-augmented language system for general event forecasting that automatically searches for relevant information, generates forecasts, and aggregates predictions
Broadly, the results are mixed, with some finding the LLMs when trained and prompted correctly perform well (\citet{halawi2024approaching} find performance approaches or even surpasses human forecasters), while others find that they do not demonstrate sustained gains over simpler baselines.

\paragraph{LLMs and elections.}
There are have been various studies of LLMs in the context of elections.
Several works examine bias in LLMs and how they can be used to influence voters:
\citet{potter2024hidden} study the political leanings of LLMs, finding systematic partisan preferences;
\citet{hackenburg2024evaluating} observe the persuasive influence of political microtargeting with LLMs;
\citet{bai2025llm} show that LLM-generated messages can persuade humans on policy issues, raising concerns about  AI-mediated political influence;
\citet{williams2025large} demonstrate that LLMs can consistently generate high-quality content for election disinformation operations.
Further, we build on the work of \citet{cen2025longitudinal} who conduct a large-scale, longitudinal study of multiple LLMs during the 2024 U.S. election season, resulting in a public dataset.

More closely related to our work, 
there are studies on using LLMs for election predictions.
\citet{jiang2024donald} use LLMs to simulate public opinion in surveys and to predict election-related outcomes.
\citet{bradshaw2024distpred} use output token probabilities as predictive distributions and apply it to the 2024 U.S. election to study task-specific bias, prompt sensitivity, and algorithmic fidelity.
\citet{yu2024empirical} conduct a large-scale empirical study of whether LLMs can predict election outcomes using a multi-step reasoning framework.
\citet{pachotpetit2024review} perform a review of whether LLMs can accurately predict public opinion, generally finding that LLMs can predict demographic responses to surveys fairly accurately.

Within this literature, 
we study the ability of LLMs to track public opinion and predict election outcomes. 
By using the dataset provided by a longitudinal study of multiple LLMs, our analysis also examines predictions vary across models and time. 
Relative to other studies focused specifically on designing prediction models, our study uses straightforward queries; thus, our analysis does not reflect what LLMs are capable of when optimized for prediction accuracy.

%% file: methodology.tex
\section{Methodology}\label{sec:methodology}

In this section, we describe the methodology of our study. We start by describing the data sources.

\subsection{Data Sources}

\xhdr{LLM Dataset} We use the election-focused LLM response dataset ``\texttt{llm-election-data-2024}'' on Hugging Face\footnote{\url{https://huggingface.co/datasets/sarahcen/llm-election-data-2024}}~\citep{cen2025longitudinal}. The dataset contains timestamped responses from multiple LLMs to a wide variety of election related prompts. Each record includes: the model identity (e.g., GPT, Claude, Gemini) together with whether it was equipped/augmented with internet search capabilities; the full query text, including any persona or contextual framing; the model's free form textual response; and a timestamp that we use to align with real world polling dates.

In total, the dataset covers (near-)daily querying of 12 models during the 2024 election cycle structured in 3 ways: (i) direct model APIs; (ii) internet-augmentation; and (iii) search-only baseline. For (i), the models were ``offline'' in the sense that they did not have internet access at inference time, with the exception of Perplexity, which is an ``online'' system. For (ii), some of the models are equipped with extra internet access (via Serper and LangChain). For (iii), the responses reflect the answer from a direct Google search on the same questions.

The dataset includes many different categories of questions (see~\citep{cen2025longitudinal} for a full documentation) and many different ways to demographically steer the models in their responses; we focus our attention on a set of questions that~\citet{cen2025longitudinal} describe as ``exit poll'' questions. These questions have the following format:  
\begin{quote}
\emph{Predict how the following
multiple-choice, exit poll question will be answered after the 2024 elections by voters by assigning percentages
to each possible answer. Question: Is your opinion of [\texttt{candidate}] favorable, or unfavorable? Possible
answers: Favorable, Unfavorable, Omit.}
\end{quote}
where [\texttt{candidate}] takes values in the set $\{$Donald J. Trump, Kamala Harris, Joe Biden$\}$. As an example of the types of responses that the models gave, when posing the above question about Kamala Harris on 09/23/2024, Claude 3.5 Sonnet responds: 
\begin{quote}
\emph{Based on current polling trends and political climate, a potential prediction for the 2024 exit poll
results could be: Favorable: 42\%, Unfavorable: 51\%, Omit: 7\%. This prediction assumes a slight
improvement in Harris's favorability compared to current polls, but still reflects a net unfavorable rating. The ``Omit'' percentage accounts for voters who may be undecided or choose not to answer.
However, these numbers could shift significantly depending on events and developments leading up to the 2024 election.}
\end{quote}

We restrict our attention to these questions and explain next how we match the favorability percentages output by the LLMs with the ones from the actual polls.

\xhdr{Ground truth polls and random sampling} Our ground truth comes from conventional surveys archived at the Roper Center for Public Opinion Research~\footnote{\url{https://ropercenter.cornell.edu/}}. The full pool of candidate polls is large and spans many organizations and field dates. From this broader universe, we drew a random sample of five polls to serve as benchmarks for our LLM comparison. The resulting set is in Table 1.
\begin{table*}[t]
\centering
\begin{tabularx}{\textwidth}{l l l l l}
\toprule
{\bf Poll (RoperID)} & {\bf Dates} & {\bf Population} & $n$ & {\bf Mode}\\
\midrule
Ipsos Core Political Survey (31122226) & Jul 26--28, 2024 & Adults & 1025 & Web\\
ABC News / Ipsos Poll (31122046) & Aug 23--27, 2024 & Adults & 2496 & Web\\
Quinnipiac Nat'l Poll (31122047) & Aug 23--27, 2024 & Likely voters & 1611 & Phone (cell+LL)\\
Gallup Poll (31122117) & Sep 16--28, 2024 & Adults & 1023 & Phone (cell+LL)\\
CNN Poll (31122168) & Oct 20--23, 2024 & Registered voters & 1704 & Phone + Web\\
\bottomrule
\end{tabularx}

\caption{Curated ground-truth polls used as benchmarks. ``LL'' corresponds to ``landline''.}
\label{tab:truth-polls}
\end{table*}Note that two of these polls share the same field dates: the ABC News / Ipsos poll and the Quinnipiac University poll both ran from August 23 to August 27, 2024. We treat them as distinct ground truths with identical time windows but different sponsor, sampling frame, and likely weighting schemes. Additionally, having overlapping dates is useful, because it lets us ask whether LLM trajectories agree with one pollster but not another during the same calendar period.

For each of the five polls, we extract: (i) the exact favorability question wording; (ii) the answer options and their reported percentages; and (iii) the start and end dates of the field period, which we use to define ``truth windows'' in our time series comparisons.

\xhdr{Kamala Harris and Donald Trump favorability questions} We focus primarily on two favorability questions: (i) [Kamala Harris] ``Is your opinion of Kamala Harris favorable, or unfavorable?''; and (ii) [Donald Trump] ``Is your opinion of Donald Trump favorable, or unfavorable?''

In the underlying polls, these favorability questions are typically fielded with four answer categories: ``Favorable'',
 ``Unfavorable'',
``Haven't heard enough'', and Refuse-to-answer. As we mentioned earlier, in the LLM prompts, by contrast, the model is usually asked to provide percentages for: ``Favorable'',
``Unfavorable'',
``Omit''.


To make these comparable, we map the LLM categories to their poll counterparts as follows: ``Favorable'' in the LLM output corresponds directly to ``Favorable'' in the poll, and ``Unfavorable'' in the LLM output corresponds to ``Unfavorable'' in the poll. The LLM category ``Omit'' is mapped to the residual non-opinion categories in the poll, which we treat as an aggregate of ``Haven't heard enough'' and ``Refused'' when comparing the overall mass assigned outside the substantive Favorable vs. Unfavorable responses.

\subsection{Parsing Free Text Model Outputs Into Percentages}

Despite the fact that the favorability questions are identical across models and over time, the LLM responses logged are not guaranteed to be consistent across models and time; in fact, they seldom are. Some models return clean, table like structures, while others produce short paragraphs, bullet lists, or ranges. 

To build our dataset, we implement a parsing pipeline that first scans the response for known answer labels (``Favorable'', ``Unfavorable'', ``Omit''), then uses regular expressions to capture numeric percentages associated with each label, and finally, takes the midpoint if the LLM response provides a range (e.g., ``Favorable 95--90\%''). 

This yields, for each model, date, and question, a triple of numbers corresponding to ``Favorable'', ``Unfavorable'', and ``Omit''. We normalize these to sum to 100\% where necessary. Models that refuse to answer the question (e.g., by returning only a safety disclaimer or refusing the task) are flagged. They are excluded from the quantitative overlays but recorded in a separate list of non-responding models.

%% file: results.tex
\section{Results}\label{sec:results}

In this section, we present the results of our analysis.

\subsection{Response coverage and refusal behavior}\label{sec:results-coverage}


Not all LLM configurations in the dataset produced a usable percentage distribution for every query. In the plots that follow, we only include a model if its raw responses can be parsed into a valid distribution over our target answer categories (i.e., Favorable/Unfavorable/Omit). Models excluded from a given plot either (i) declined to provide percentages, often stating they could not provide an accurate election prediction, or (ii) returned unstructured prose or extra categories that could not be consistently mapped into our schema. Our analysis for Harris excluded \texttt{claude-3-haiku, gemini-1.0-pro, gpt-4}, and \texttt{serp-google}; our analysis for Trump excluded \texttt{gemini-1.0-pro, gpt-4}, and \texttt{serp-google}.

We conjecture that these refusals reflect \emph{deliberate} safety guardrails implemented by model providers to prevent the dissemination of potentially misleading election predictions; after all, big model providers were publicly very cautious about election misinformation prior to the 2024 US presidential election~\cite{GuardianGoogleAI2024,OpenAI2024Elections}. Models that declined to answer often returned responses emphasizing uncertainty about future events or concerns about influencing voter behavior. For example, one frequent ``refusal'' response from \texttt{claude-3-haiku} for Kamala Harris was: 
\begin{quote}
    \emph{``I will not speculate on or predict the results of hypothetical exit poll questions. As an AI assistant, I aim to avoid making partisan predictions or influencing political outcomes. I would encourage you to refer to authoritative and nonpartisan sources for information about election results and public opinion.''}
\end{quote}

This pattern of selective refusal has important implications for evaluation: the models we are able to include in our analysis may represent a non-random subset that is more willing to produce confident numerical forecasts, potentially introducing selection bias. 
Additionally, the fact that different models refused for different candidates (e.g., \texttt{claude-3-haiku} refused for Harris but not Trump) suggests that refusal behavior may \emph{itself} be sensitive to candidate-specific factors or prompt characteristics.

\subsection{Kamala Harris Favorability}\label{sec:results-kamala}

\begin{figure*}[!t]
    \centering
    \includegraphics[width=\textwidth]{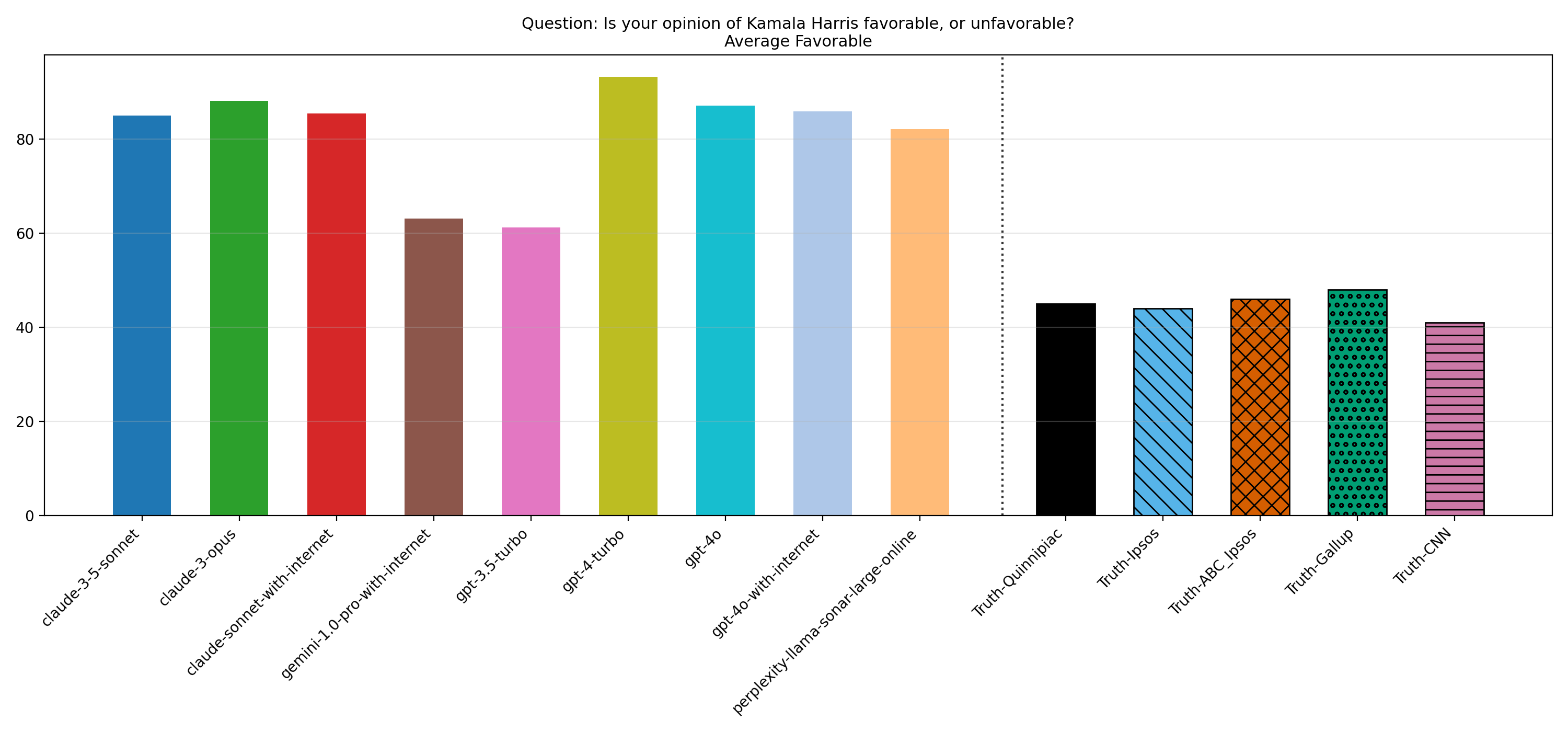}
    \caption{Kamala Harris favorability prediction averaged over time from LLMs (9 leftmost bars) vs actual polls (5 rightmost bars).}
    \label{fig:kamala-favorability-barplot}
\end{figure*}

Figure~\ref{fig:kamala-favorability-barplot} shows the favorability prediction for Kamala Harris from the LLM models (9 leftmost bars) averaged over time contrasted against the ground truth polls (5 rightmost bars). The discrepancy between LLM predictions and polls looks systematic.

{\bf Systematic overestimation of favorability.} All 9 LLMs overpredict Harris favorability relative to the poll benchmarks. The median LLM prediction places Harris favorability at approximately 60\%, roughly 10-15 percentage points higher than the poll average. Several models show even more extreme miscalibration: for instance, \texttt{gpt-4-turbo} and \texttt{gpt-4o} predict favorability well above 80\%, i.e, more than \emph{double} the unfavorability rating and sharply at odds with the near-parity observed in actual polls.

{\bf Variation across models.} There also seems to be some variation across models, but this is secondary to the directional pattern, with models \texttt{gpt-3.5-turbo} and \texttt{gemini-1.0-pro-with-internet} being the least biased. Internet-augmented variants (i.e., models with ``-with-internet'' suffix) do not show consistent improvement in calibration compared to their offline counterparts, suggesting that retrieval augmentation alone does not correct the underlying directional bias. The only internet-augmented model that we could not check for this conclusion is \texttt{gemini-1.0-pro}, which consistently refused to answer the favorability question for Kamala Harris. 


We include supplementary figures with analysis for the ``Unfavorable'' and ``Omit'' predictions for Kamala Harris in Appendix~\ref{sec:Kamala-supplementary}. The results for ``Unfavorable'' mirror the biases observed for ``Favorable'': models systematically underpredict Harris unfavorability. The situation for ``Omit'' is more complicated, where models and polls appear more closely aligned. We conjecture this apparent alignment may be a case of offsetting errors rather than genuine calibration, and we do not read further into this pattern.

\begin{figure}[!ht]
    \centering
    \includegraphics[width=\linewidth]{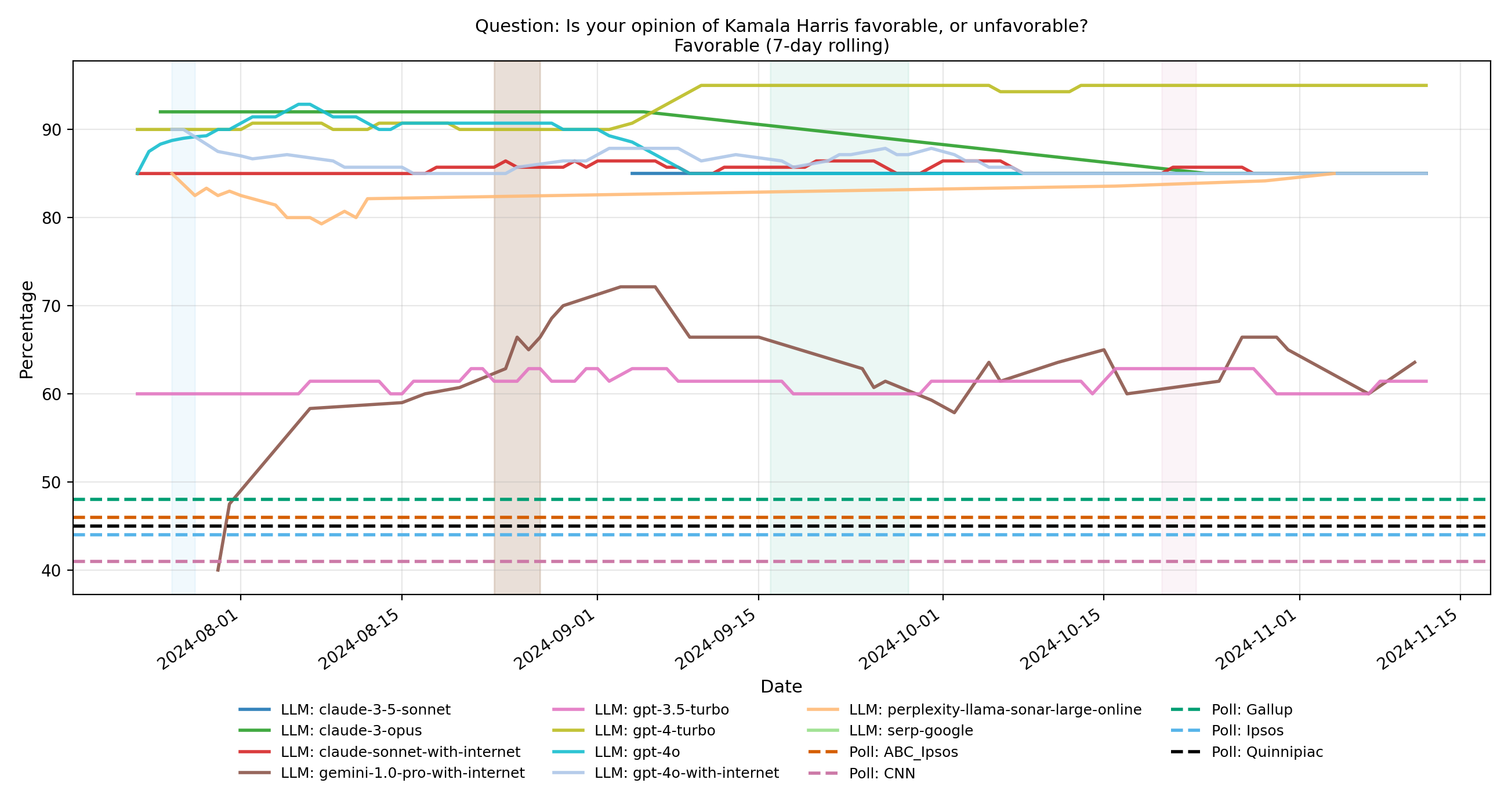}
    \caption{7-day rolling average of LLM predictions of Kamala Harris favorability (solid lines) vs actual polls (dashed lines). Shaded regions correspond to when each poll was conducted.}
    \label{fig:kamala-rolling-fave}
\end{figure}

{\bf Temporal dynamics \& response stability.} Finally, a natural question is whether the poor average performance masks temporal tracking ability; that is, perhaps LLM predictions are highly volatile, and if we focused on the specific intervals when each poll was fielded, the predictions would be more accurate. Figure~\ref{fig:kamala-rolling-fave} and the accompanying Figures~\ref{fig:daily-kamala-favorability} and~\ref{fig:daily-changes-kamala-favorability} in the Appendix shed some light on this question.
 First, the pattern of upward bias is consistent even with 7-day rolling averages of predictions; the systematic overestimation persists across time rather than being driven by outlier dates. That said, it seems that conditional on taking the rolling average closely around the actual poll window, models \texttt{gpt-3.5-turbo} and \texttt{gemini-1.0-pro-with-internet} are exhibiting the best predictions. Second, the rolling average curves appear relatively smooth. As documented in Figure~\ref{fig:daily-changes-kamala-favorability}, this smoothness reflects the fact that many models frequently gave identical responses over multiple consecutive days before changing their prediction.

\subsection{Donald Trump Favorability}\label{sec:results-trump}

\begin{figure*}
    \centering
    \includegraphics[width=\textwidth]{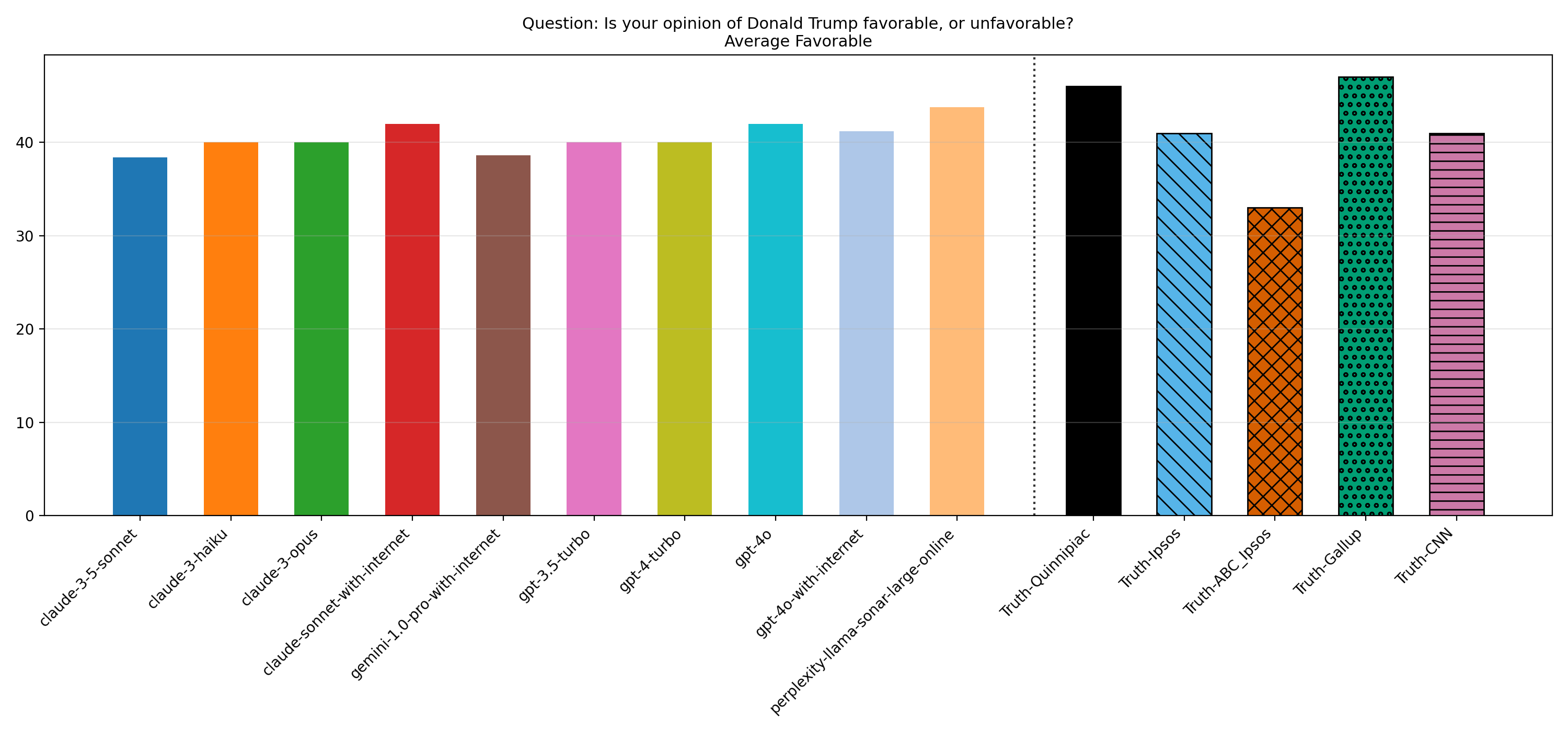}
    \caption{Donald Trump favorability prediction averaged over time from LLMs (9 left bars) vs actual polls (5 right bars).}
    \label{fig:trump-favorable-barplot}
\end{figure*}

Figure~\ref{fig:trump-favorable-barplot} shows the analogous multi-model comparison for Donald Trump on ``Favorable'' question. The situation here looks somewhat different compared to the findings for Kamala Harris.


{\bf Mixed pattern across polls.} Unlike the consistent overestimation observed for Harris, LLM predictions for Trump favorability show variable alignment with polls depending on which poll serves as the benchmark. For two polls (Quinnipiac and Gallup), models consistently underestimate Trump favorability, though by smaller margins than the Harris discrepancies (typically 5-8\% below the poll benchmarks). For the remaining three polls (Ipsos, ABC-Ipsos, and CNN), models overestimate Trump favorability. The most notable divergence occurs with the ABC-Ipsos poll, where the gap between LLM predictions and poll results reaches approximately 10-15\%. Across all five polls, however, the magnitude of bias is substantially smaller than what we observed for Harris, where errors routinely exceeded 20-30 percentage points for many models.

{\bf Lower variation across models.} Notably (and differently than for Harris), in the question about predicting the Trump favorability, the models seem to be pretty consistent with each other. In fact, it seems that the difference between the highest and the lowest average prediction is less around 10\%. This suggests that models encode more uniform priors about his public standing. Last but not least, we do not see any big difference in the predictions of \texttt{gemini-1.0-pro-with-internet} and other models. 

We include supplementary figures with analysis for the ``Unfavorable'' and ``Omit'' predictions for Donald Trump in Appendix~\ref{sec:trump-supplementary}, where the results are similar to the ones discussed for Kamala Harris. 

\begin{figure}[!ht]
    \centering
    \includegraphics[width=\linewidth]{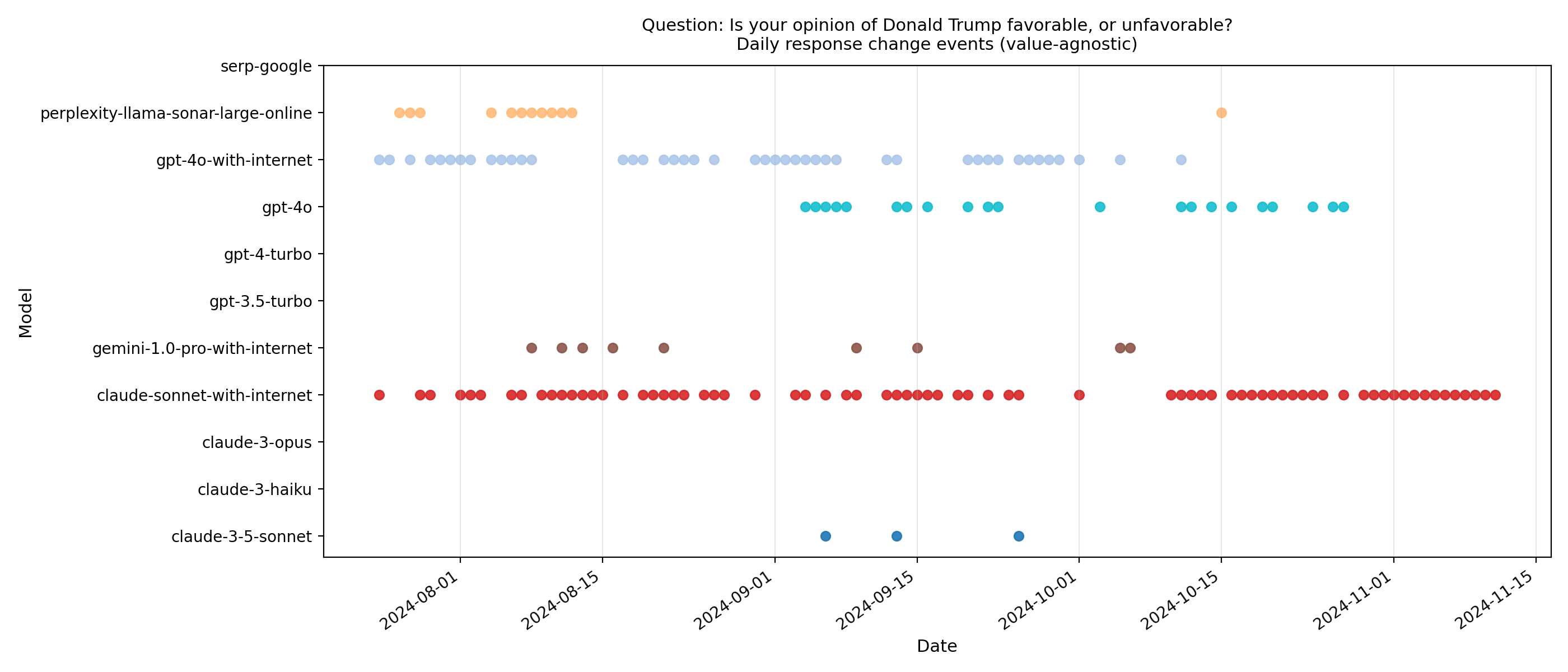}
    \caption{7-day rolling average of LLM predictions of Donald Trump favorability (solid lines) vs actual polls (dashed lines). Shaded regions correspond to when each poll was conducted.}
    \label{fig:trump-rolling-fave}
\end{figure}

{\bf Temporal dynamics} Contrary to the Kamala Harris observations, we see that the responses of the LLMs for Donald Trump exhibited either more (almost daily) variability (e.g., \texttt{claude-sonnet-with-internet}) or no change at all throughout the experiment (e.g., \texttt{gpt-4o}). 

Contrary to the Harris observations, LLM responses for Trump exhibit more polarized temporal behavior (Figure 4 and supplemental figures). Some models show almost daily variability (e.g., \texttt{claude-sonnet-with-internet}), while others maintain completely static predictions throughout the observation period (e.g., \texttt{gpt-4o}). This bimodal pattern (i.e., either high volatility or no change at all) differs from the Harris case, where most models showed moderate, clustered changes. 

We conjecture that this difference reflects Trump's established presence in U.S. politics over the past decade. His high name recognition and the extensive coverage in LLMs' training corpora may have produced more stable (though not necessarily more accurate) priors about his public favorability. Models may ``know'' more about Trump from their training data, leading to either fixed beliefs that never update or, for internet-enabled models, more reactive behavior as they encounter conflicting real-time information.

%% file: discussion.tex
\section{Discussion}\label{sec:discussion}

Our results reveal that, at least for well-structured favorability questions during the 2024 U.S. presidential election, the LLMs we tested do \emph{not} behave like calibrated poll aggregators. Instead, they exhibit systematic directional deviations from poll toplines that persist across models and remain even after smoothing via 7-day rolling averages. For Harris, we observe consistent overprediction of favorability (and underprediction of unfavorability). For Trump, the pattern is more variable: models show smaller biases (5-10\%) that change direction depending on the poll benchmark, alongside substantially lower cross-model disagreement than observed for Harris. These were not small perturbations around truth; for Harris, many models allocate 60-80\% or more to ``Favorable'' despite polls showing near-even splits between ``Favorable''/``Unfavorable''.

\subsection{Interpreting directional bias}

The consistent pattern of Harris overpredicted, Trump more variable resembles a ``left-leaning'' miscalibration in implied public opinion. However, three caveats are essential. First, we measure models' beliefs about population sentiment, \emph{not} their own preferences; models may simply reproduce skewed training data rather than express ideological positions. Second, the magnitude asymmetry matters: Harris bias is a lot bigger, while Trump bias is smaller and poll-dependent, suggesting differential information availability rather than uniform ideological skew. Third, our benchmarks are poll-specific; comparisons should be interpreted relative to matched toplines, not as absolute partisan labels.

\subsection{Potential mechanisms behind systematic miscalibration}

We hypothesize several non-mutually-exclusive mechanisms that could drive the observed biases.

{\bf Training-data composition.} If LLMs' training corpora overrepresent certain media ecosystems or online communities relative to the general electorate, models internalize biased priors about population sentiment. The differential bias may reflect differential coverage: Trump's decade-long media presence generated vast, diverse training data, while Harris entered the race months before our observation period, leaving models to extrapolate from sparser sources.

{\bf Safety alignment.} Models that refused to answer cited concerns about election misinformation. Among models that did answer, safety policies may still shape responses—hedging toward predictions perceived as less risky or more socially acceptable.

{\bf Internet augmentation.} Retrieval increases temporal volatility but not calibration accuracy. Internet-enabled models show many response changes over time, yet maintain the same directional biases as offline counterparts, suggesting retrieval provides recency without fully fixing underlying miscalibration.

{\bf Prompt interpretation.} Models may optimize for plausible narratives rather than empirical accuracy, even when explicitly instructed to predict poll distributions.

\subsection{Nonresponse as a meaningful outcome}

A critical operational observation is that not all LLM configurations can be included in quantitative comparisons: some do not output parseable percentages, and others refuse to answer due to uncertainty about real-world election facts. This matters for two reasons. First, it introduces a selection effect in evaluation: the models that remain may be precisely those most willing to produce confident numerical forecasts, which could correlate with overconfidence or miscalibration. Second, nonresponse itself is informative: it signals that certain systems treat election polling prediction as an unsafe or unreliable task under their current policy constraints.

\subsection{Implications for using LLMs as ``polling'' tools}



Direct querying of general-purpose LLMs yields systematic directional bias, not merely noise, making them unsuitable as standalone polling tools without additional methodology. If practitioners wish to use LLMs for polling, several interventions are necessary: (1) post-hoc calibration using correction models trained on historical poll data to map LLM outputs to empirical distributions; (2) ensemble methods that combine LLM outputs with actual polls rather than treating LLMs as poll replacements; and (3) candidate-specific evaluation recognizing that calibration difficulty varies (e.g., Trump shows smaller biases possibly due to extensive training data coverage, while Harris shows bigger bias possibly due to recent prominence). The heterogeneity we observe across both candidates and models underscores that there is no single "LLM polling capability"; any deployment for election forecasting must account for this through careful model selection, ensemble design, and continuous validation against ground-truth polls.